\documentclass[journal=jacsat,manuscript=article]{achemso}
\setkeys{acs}{usetitle=true}
\usepackage{xcolor}
\usepackage{graphicx}
\usepackage{amsmath}
\usepackage{caption}
\usepackage{subcaption}
\usepackage{hyperref}
\usepackage[version=3]{mhchem}
\usepackage{xr}
\usepackage[T1]{fontenc}  
\usepackage{float} 
\usepackage{makecell} 
\usepackage{setspace} 
\usepackage{xr-hyper}

\makeatletter
\newcommand*{\addFileDependency}[1]{
  \typeout{(#1)}
  \@addtofilelist{#1}
  \IfFileExists{#1}{}{\typeout{No file #1.}}
}
\makeatother

\newcommand*{\myexternaldocument}[1]{%
    \externaldocument{#1}%
    \addFileDependency{#1.tex}%
    \addFileDependency{#1.aux}%
}
\myexternaldocument{si}
\newcommand{\e}[1]{\ensuremath{\times 10^{#1}}} 

\title{Robust Machine Learning Inference from X-ray Absorption Near Edge Spectra through Featurization}

\author{Yiming Chen}
\affiliation[UCSD]{Department of NanoEngineering, University of California San Diego, 9500 Gilman Dr, La Jolla, CA 92093, United States}
\alsoaffiliation[CNM]{Center for Nanoscale Materials, Argonne National Laboratory, 9700 South Cass Avenue, Lemont, IL 60439, United States}

\author{Chi Chen}
\affiliation[UCSD]{Department of NanoEngineering, University of California San Diego, 9500 Gilman Dr, La Jolla, CA 92093, United States}

\author{Inhui Hwang}
\affiliation[APS]{Advanced Photon Source, Argonne National Laboratory, 9700 South Cass Avenue, Lemont, IL 60439, United States} 

\author{Michael J. Davis}
\affiliation[CSE]{Division of Chemical Sciences and Engineering, Argonne National Laboratory, 9700 South Cass Avenue, Lemont, IL 60439, United States} 

\author{Wanli Yang}
\affiliation[ALS]{Advanced Light Source, Lawrence Berkeley National Laboratory, 6 Cyclotron Rd, Berkeley, CA 94720, United States}

\author{Chengjun Sun}
\affiliation[APS]{Advanced Photon Source, Argonne National Laboratory, 9700 South Cass Avenue, Lemont, IL 60439, United States} 

\author{Gihyeok Lee}
\affiliation[ALS]{Advanced Light Source, Lawrence Berkeley National Laboratory, 6 Cyclotron Rd, Berkeley, CA 94720, United States}

\author{Dylan McReynolds}
\affiliation[ALS]{Advanced Light Source, Lawrence Berkeley National Laboratory, 6 Cyclotron Rd, Berkeley, CA 94720, United States}

\author{Daniel Allan}
\affiliation[NSLS-II]{National Synchrotron Light Source II, Brookhaven National Laboratory, Upton, NY 11973, United States}

\author{Juan Marulanda Arias}
\affiliation[NSLS-II]{National Synchrotron Light Source II, Brookhaven National Laboratory, Upton, NY 11973, United States}

\author{Shyue Ping Ong}
\email{ongsp@ucsd.edu}
\affiliation[UCSD]{Department of NanoEngineering, University of California San Diego, 9500 Gilman Dr, La Jolla, CA 92093, United States}

\author{Maria K.Y. Chan}
\email{mchan@anl.gov}
\affiliation[CNM]{Center for Nanoscale Materials, Argonne National Laboratory, 9700 South Cass Avenue, Lemont, IL 60439, United States}

\begin{document}
\maketitle

\begin{abstract}
\item X-ray absorption spectroscopy (XAS) is a commonly-employed technique for characterizing functional materials. In particular, x-ray absorption near edge spectra (XANES) encodes local coordination and electronic information and machine learning approaches to extract this information is of significant interest. To date, most ML approaches for XANES have primarily focused on using the raw spectral intensities as input, overlooking the potential benefits of incorporating spectral transformations and dimensionality reduction techniques into ML predictions. In this work, we focused on systematically comparing the impact of different featurization methods on the performance of ML models for XAS analysis. We evaluated the classification and regression capabilities of these models on computed datasets and validated their performance on previously unseen experimental datasets. Our analysis revealed an intriguing discovery: the cumulative distribution function (CDF) feature achieves both high prediction accuracy and exceptional transferability. This remarkably robust performance can be attributed to its tolerance to horizontal shifts in spectra, which is crucial when validating models using experimental data. While this work exclusively focuses on XANES analysis, we anticipate that the methodology presented here will hold promise as a versatile asset to the broader spectroscopy community. 
\end{abstract}

\section{Introduction}
X-ray absorption spectroscopy (XAS) is a versatile characterization technique to probe the oxidation states\cite{ravelSimultaneousXAFSMeasurements2010}, spin states\cite{boillotPressureinducedSpinstateCrossovers2002}, and coordination environment\cite{hudson-edwardsOriginFateVanadium2019} in materials. A typical XAS spectrum can be divided into two regions depending on the energy range. The X-ray absorption near-edge structure (XANES), the region within 50 eV of absorption onset, is more sensitive to the oxidation states and coordination environments. The extended X-ray absorption fine structure (EXAFS), on the other hand, encodes information about the neighboring atoms and excited states. \cite{torrisiRandomForestMachine2020} While the quantitative analysis of EXAFS is relatively well-established with explicit equations to fit computed spectra, the analysis for the XANES is still constrained by the limited number of reference spectra \cite{choudharyRecentAdvancesApplications2022,newvilleFundamentalsXAFS2014}. 

Nevertheless, advances in operando measurements, computational approaches\cite{rehrTheoreticalApproachesXray2000,grootMultipletEffectsXray2005a,laskowskiUnderstandingXrayAbsorption2010} and computing power in recent years have greatly alleviated such data scarcity. For example, in previous studies,\cite{mathewHighthroughputComputationalXray2018b,chenDatabaseInitioLedge2021} some of the present authors have developed the XASDB, the world's largest database of computed XANES hosting approximately 500,000 K-edge and 140,000 L-edge spectra. Such large, computational XAS databases provide a highly useful complement to experimental XAS data. While experimental XAS data collection typically focuses on limited chemistries (usually a single phase or phase mixture) under a wide variety of conditions, computational databases such as the XASDB provide spectra on a broad diversity of structures and chemistries, albeit under limited conditions (typically 0K structures from density functional theory (DFT) calculations).

With the increasing availability of experimental and computed XANES data, there have been significant research efforts into the application of machine learning (ML) techniques to extract insights and make predictions from XAS. For instance, Guda et al. applied multivariate curve resolution methods on operando XANES spectra to isolate individual species/phases from the multicomponent data mixture in a catalyst system\cite{gudaQuantitativeStructuralDetermination2019}. \textcolor{black}{Supervised machine learning techniques have been widely applied to establish a correspondence between spectra and target properties, either in a forward or inverse direction\cite{unruh_theoryaiml_2022}.} A large body of work focused on the accurate inference of target properties from XANES. \textcolor{black}{For example, electronic structures such as oxidation states and electronic configuration of $d$ states can be determined from K and L-edge XANES spectra through ML-based approaches.\cite{miyazatoAutomaticOxidationThreshold2019,luder_determining_2021} }
The determination of coordination environment has also been investigated in several studies\cite{liDeepLearningModel2019,carboneClassificationLocalChemical2019,zhengRandomForestModels2020}. Despite using different coordination environment descriptors, remarkably high accuracy of above 80\% has been demonstrated in all cases. Another study also reported a >80\% accuracy in determining whether an inorganic material is topological from the XANES data. \textcolor{black}{In addition to inverse property determination from spectra, researchers also explored how to perform forward modeling of XAS spectra based on structural information using neural networks to achieve quantitative accuracy and derive uncertainty matrix. \cite{carbone_machine-learning_2020,david_towards_2023} Beyond XAS, other techniques such as X-ray photoelectron spectroscopy (XPS) and phonon density of states (DoS) prediction were made possible through machine learning approaches.\cite{golzeAccurateComputationalPrediction2022,chen_direct_2021} }These studies exemplify what ML models are capable of in terms of both accuracy and generalizability. 

The majority of the ML models in XAS analysis have thus far utilized the raw spectra, i.e., the paired values of energies and intensities, as the input features. Only a few studies have investigated how the ML models could benefit from additional transformations and dimensionality reduction of the raw spectra. One such example is that of Torrisi et al., who found that polynomial-fitted features from XAS could aid model interpretability by incorporating local trends and focusing on spectral shape that was buried under individual intensity points\cite{torrisiRandomForestMachine2020}. Similarly, Tetef et al. showed that a t-distributed stochastic neighbor embedding (t-SNE) of XANES not only achieves superior prediction accuracy in classifying aromaticity, but is also able to distinguish the finer sub-classes for sulforganics\cite{tetefUnsupervisedMachineLearning2021}. Latent representation was also explored to generate a low-dimensional representation of Pd K-edge XANES spectra that could maintain the spectrum-structure relationship and provide an innovative pathway to identify the key factors for spectral changes\cite{routhLatentRepresentationLearning2021}. 

In this work, we investigated different featurization approaches and benchmarked their effect on the classification and regression performance of various ML models for XAS analysis. We have selected the Ni K-edge XANES of \ce{Li_zNi_{x}Mn_{y}Co_{1-x-y}O2} (NMC) as our system of interest. NMC is a family of cathodes of major importance in rechargeable Li-ion batteries because of their high energy density and long-term cyclability.\cite{manthiramReflectionLithiumionBattery2020} 
During the cycling of a battery where Li is extracted or inserted, the transition metals, in particular, Ni, undergo oxidation state changes and with it, corresponding bond length changes.\cite{chakrabortyLayeredCathodeMaterials2020} The K-edge XANES from operando or ex-situ XAS experiments is often used to track such changes in NMC cathodes during, or at different stages of, battery cycling. 
Evaluating different ML models and featurization approaches based on computed data, we find that ensemble tree-based methods such as gradient boosting or random forest models tend to outperform other ML models in terms of predicting bond length regression and oxidation state classification, in line with the findings of previous studies.\cite{zhengRandomForestModels2020,torrisiRandomForestMachine2020,gudaQuantitativeStructuralDetermination2019} Several featurization approaches result in similarly accurate inference on computed XANES, such as cumulative distribution function (CDF), peak feature and continuous wavelet transform (CWT). 
However, when performing inference on experimental data, we find that the performance of tree-based models can be enhanced by performing a cumulative distribution function (CDF) transformation on the XAS. 

\section{Methods}
\subsection{Overall workflow}

\begin{figure}[H]
	\begin{center}
		\includegraphics[width=0.85\textwidth]{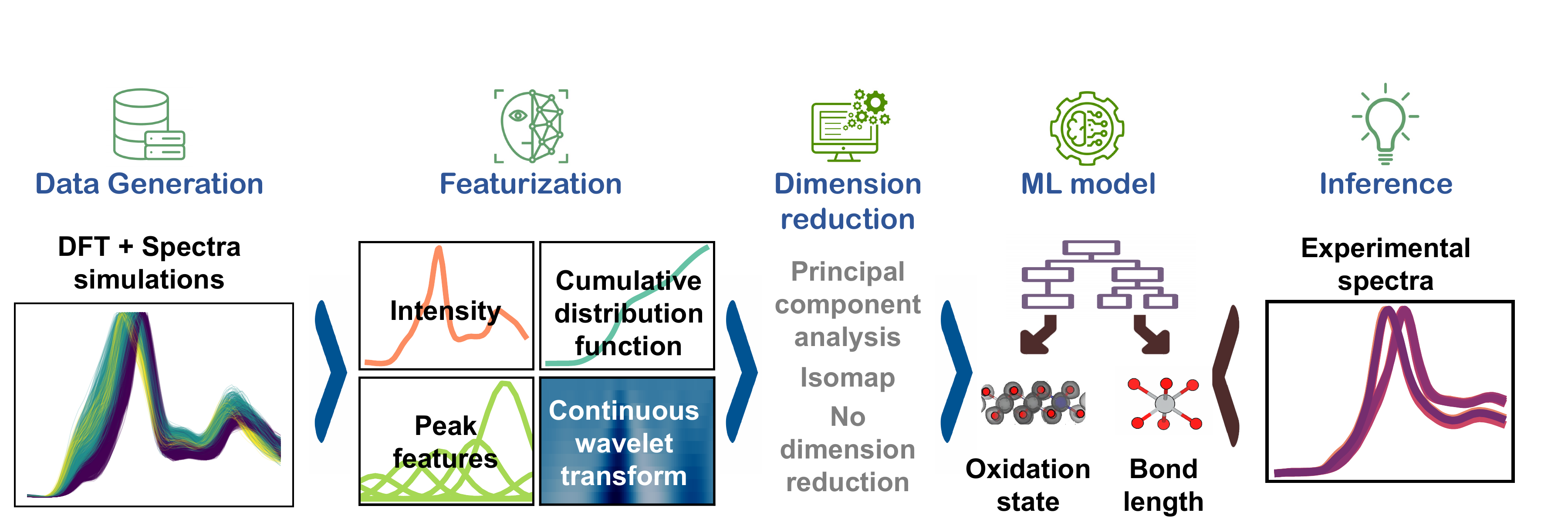}
		\caption{
			\label{fig:workflow}
			Schematic diagram of benchmarking feature space for supervised machine learning.}
	\end{center}
\end{figure}

\begin{figure}[H]
	\begin{center}
		\includegraphics[width=1\textwidth]{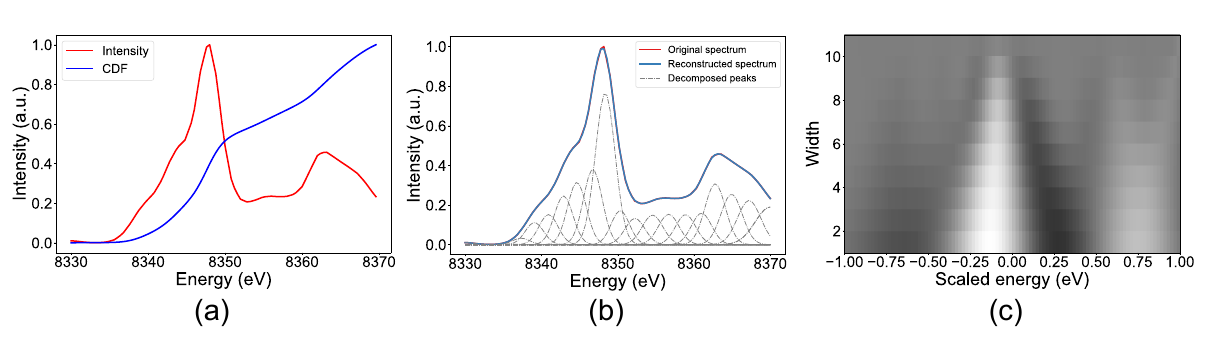}
		\caption{
			\label{fig:spec_transformation}
			Visualization of spectral transformation for (a)intensity and CDF, (b) peak feature and (c) CWT. }
	\end{center}
\end{figure}

Figure \ref{fig:workflow} shows the overall workflow for this work. The raw XANES were first interpolated and projected onto the same energy scale with 100 individual intensities, which were then normalized to the maximum intensity. The initial feature vectors were then derived by performing four different transformations (see Figure \ref{fig:spec_transformation}):

\begin{enumerate}
    \item Original intensity: The original vector of 100 intensities for each spectrum was used. 
    \item Cumulative distribution function (CDF): The CDF of each spectrum was computed and normalized to the maximum value.
    \item Peak feature: Each spectrum was decomposed into a sum of 20 Gaussian peaks using non-linear least square fitting in scikit-learn\cite{pedregosaScikitlearnMachineLearning2011}. The information for each Gaussian peak was simplified to three parameters: peak center, peak amplitude and peak width. During fitting, each peak center was limited to vary within a 5 eV range and each peak width, FWHM, was bounded between 0 and 6 eV. For each decomposed spectrum, a total of 20 Gaussian peaks are ordered by their peak energies in an increasing order and converted into a vector containing 60 values in the form of [\ce{center_{1}},\ce{amplitude_{1}},\ce{width_{1}} ... \ce{center_{20}},\ce{amplitude_{20}},\ce{width_{20}}]. 
    \item Continuous wavelet transformation (CWT): The CWT is a common transformation used in EXAFS analysis.\cite{funkeNewFEFFBased2007,munozContinuousCauchyWavelet2003,timoshenkoWaveletDataAnalysis2009} However, relatively few studies have explored its application in XANES analysis. A Ricker wavelet function and widths of integers from 1 to 10 were applied for CWT transformation. The resultant CWT transformation is a 2d array with dimension (10, 100).
\end{enumerate}

For each featurization approach, we then optionally applied three dimensionality reduction techniques - principal component analysis (PCA), Isomap, or Autoencoder (AE). PCA\cite{hotellingANALYSISCOMPLEXSTATISTICAL1933,pearsonLIIILinesPlanes1901a} is a linear decomposition technique that has been applied widely in the spectroscopy field for data analysis\cite{tetefUnsupervisedMachineLearning2021,chenMachineLearningNeutron2021}. Similarly, Isomap\cite{tenenbaumGlobalGeometricFramework2000} is a non-linear isometric mapping dimensionality reduction technique to capture the underlying geometry of a data manifold. Autoencoder aims to learn how to efficiently compress and encode data so that it can reconstruct the data from the latent space representation in a way as close to the original input as possible. Tensorflow\cite{abadiTensorFlowSystemLargescale} was used to build AE models. To allow a fair comparison between different dimensionality reduction techniques, all features were reduced to 3-dimensional ones. \textcolor{black}{The python code to featurize the spectra, example inputs, trained models and computational dataset are open to the public on Github: \href{https:github.com/MaterialEyes/FeatureXAS}{github.com/MaterialEyes/FeatureXAS}.} \textcolor{black}{The dataset is also hosted on AIMMDB: \href{https://aimm.lbl.gov/ui/browse/nmc_sim_vasp} {aimm.lbl.gov/ui/browse/nmc\_sim\_vasp} and  Zenodo: \href{https://zenodo.org/records/10476278}{zenodo.org/records/10476278}.}

\subsection{Machine learning models}
All the ML models used in this work were implemented using the scikit-learn package\cite{pedregosaScikitlearnMachineLearning2011}. Four supervised ML models - ridge regression, gradient boosting (GB), random forest (RF), and multi-layer perceptron (MLP) - were assessed in terms of the performance in inferring oxidation states and bond length from XANES. In addition, we define a ``baseline'' (dummy - DUM) performance as the ratio of most abundant category for classification or the error based on always predicting the mean of dataset for regression. 

\subsection{Machine learning targets}

We selected two targets for machine learning in this work. The first regression target is the average Ni-O bond length $\bar{L}$ of each \ce{NiO6} octahedra, i.e., 
\begin{equation}
    \bar{L} = \frac{\sum_{i=1}^6 L_i}{6}
\end{equation}
where $L_i$ is the length of the $i$th bond in an \ce{NiO6} octahedra.

The second classification target is the oxidation state of Ni, which is determined from the integrated DFT spin density. \textcolor{black}{Integrated spin density measures the magnetic net moment up to a certain radius (e.g., 2Å in this study) around the atom and this method has been widely used in computational materials science to determine the oxidation states of transition metal elements\cite{liuElucidatingLimitLi2019,walshElectronCountingSolids2017,fengStudiesFunctionalDefects2019}. In this specific case, \ce{Ni^{2+}}, \ce{Ni^{3+}} and \ce{Ni^{4+}} have electronic configurations of \ce{t_{2g}^{6}e_{g}^{2}}, \ce{t_{2g}^{6}e_{g}^{1}} and \ce{t_{2g}^{6}e_{g}^{0}}, respectively and those configurations represent 2, 1 and 0 net moment. } 

\subsection{Dataset}

\begin{figure}[H]
			\begin{center}	\includegraphics[width=0.85\textwidth]{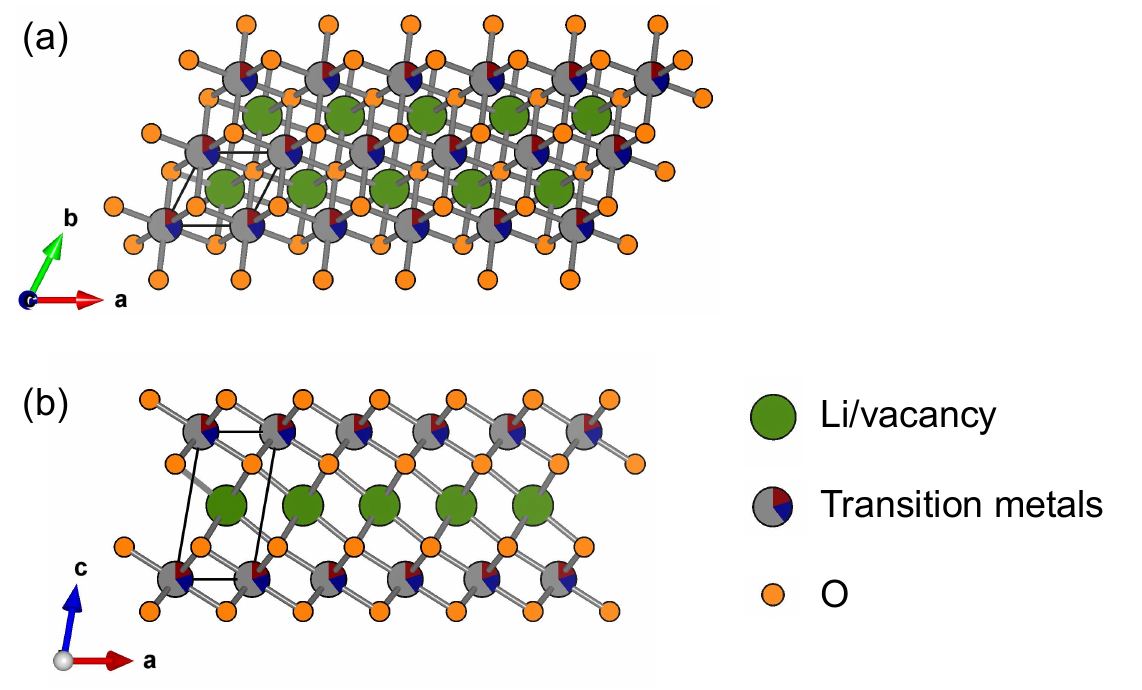}
				\caption{
					\label{fig:NMC_crystal}
					Crystal structure for the NMC supercell along (a)c-axis and (b)b-axis. Each supercell contains 10 formula units of \ce{Li_zNi_{x}Mn_{y}Co_{1-x-y}O2}.}
			\end{center}
		\end{figure}

The computed XAS dataset was obtained by performing density functional theory (DFT) calculations on NMC structures. An NMC material is typically abbreviated based on the relative ratios of transition metals in its formula. For example, NMC811 and NMC111, two common cathode compositions, refer to \ce{LiNi_{0.8}Mn_{0.1}Co_{0.1}O2} and \ce{LiNi_{1/3}Mn_{1/3}Co_{1/3}O2}, respectively. The supercell used in this work is a $5\sqrt{3} \times 2\sqrt{3}\times 1/3$ cell of the \ce{LiCoO2} conventional cell (structure prototype: $\alpha$-\ce{NaFeO2}, space group: R$\bar{3}$m), as shown in Figure \ref{fig:NMC_crystal}. The Co is then replaced with an appropriate mixed occupancy of Ni:Mn:Co to obtain the NMC622, NMC811 and NMC721 compositions and the Li site is replaced with a partial occupancy to obtain lithiation levels in intervals of 0.1. An enumeration is then carried out to obtain all symmetrically-distinct orderings of transition metals and Li/vacancy. The detailed structure distribution can be found in Figure \ref{fig:structure_distribution}. All structures were fully relaxed before XAS computations (see below). A total of 2831 site-specific Ni K-edge XANES spectra were obtained for around 700 NMC structures. 
To account for the offset in energies between VASP-computed and experimental NMC spectra, all computed spectra were shifted higher in energy by a constant 120 eV. The spectra were also normalized such that the maximum peak intensity has a value of 1. One-dimensional interpolation was applied to ensure the same energy grid (i.e., 0.4 eV increment) and energy range (i.e, 8330-8370 eV) for all spectra in the dataset. Gaussian broadening with a full width half maximum (FWHM) of 3 eV was applied on the raw spectra to mimic the instrumental broadening. The whole site-wise spectra dataset was split into a train and a test dataset with an 80:20 ratio. 

The experimental NMC datasets comprise ex-situ Ni K-edge XANES measurements taken for NMC622 and NMC721 batteries. The incident X-ray energy was selected using a three-quarters-tuned Si(111) double crystal monochromator and Harmonic rejection mirror at the 20-BM beamline of the Advanced Photon Source (APS). The XAFS data processing was performed using the Demeter package, a software package for the analysis of X-ray absorption spectroscopy, following standard analysis procedures\cite{ravelATHENAARTEMISHEPHAESTUS2005}.
\textcolor{black}{We also collected two independent datasets for two Li-rich systems, 0.5\ce{Li_{2}MnO_{3}}$\cdot$0.5\ce{LiMn_{0.5}Ni_{0.35}Co_{0.15}O_{3}}\cite{atesHighRateLirich2015} and \ce{Li_{1.2}Mn_{0.6}Ni_{0.2}O_{2}}\cite{li_improving_2022}, using WebPlotDigitizer\cite{Rohatgi2022}.}
To compare with the experimental spectra, we constructed three site-averaged datasets from site-wise spectra. The first dataset contained only site-averaged spectra for NMC622 structures. The second dataset was for the NMC721 system, while the third one contained all site-averaged spectra for NMC622, NMC721, and NMC811. These datasets comprised 181, 115, and 409 spectra, respectively.

\subsection{Density functional theory calculations}

All spin-polarized DFT calculations were performed using version 6.1 of the Vienna Ab initio Simulation package (VASP)\cite{kresseEfficientIterativeSchemes1996} within the projector-augmented wave approach \cite{blochlProjectorAugmentedwaveMethod1994}. The exchange-correlation functional used for structural relaxation and spectroscopy calculation was strongly-constrained and appropriately normed (SCAN) functional \cite{sunStronglyConstrainedAppropriately2015a}. Hubbard U values of 2.43 eV, 2.93 eV, and 2.86 eV were applied for the $d$ orbitals of Ni, Mn and Co, respectively, based on previous work by Wang et al.\cite{wangPredictingAqueousStability2020a} \textcolor{black}{who developed Hubbard U correction for SCAN functional using regular PAW potentials}. All calculations were initialized in a ferromagnetic configuration with Co in a low spin state and Ni and Mn in a high spin state, consistent with previous studies\cite{muellerEvaluationTavoriteStructuredCathode2011,tangProbingSolidSolid2018,guoDesignPrinciplesAqueous2020}. The plane wave energy cutoff was set to 450 eV and k-points density was 1500/ (\# of atoms), which is similar to parameters used in Materials Project\cite{jainCommentaryMaterialsProject2013}. The energy and force convergence criteria were 1\e{-4} eV and -0.05 eV \r{A}$^{-1}$, respectively. All input generation, and output analysis were performed using the open-source Python Materials Genomics (pymatgen) package\cite{ongPythonMaterialsGenomics2013}.

VASP6 was also used to compute the Ni K-edge XAS using the super-cell core-hole method\cite{karsaiEffectsElectronphononCoupling2018} \textcolor{black}{that creates a 1s hole and places that electron into the conduction bands.}.  For spectroscopy calculations, the GW PAW potential, which is a harder pseudopotential that includes more electronic states, was used instead of the typical PAW ones used for structural relaxations. In addition to VASP, we also performed XAS calculations using three well-established computational codes - FEFF\cite{rehrParameterfreeCalculationsXray2010b}, FDMNES\cite{bunauSelfconsistentAspectsXray2009}, and OCEAN\cite{vinsonBetheSalpeterEquationCalculations2011} - for benchmarking purposes. A detailed discussion of computational XAS theory can be found in several excellent reviews\cite{grootMultipletEffectsXray2005a,rehrTheoreticalApproachesXray2000,laskowskiUnderstandingXrayAbsorption2010}. Briefly, FEFF and FDMNES employ real-space full multiple-scattering theory within the muffin-tin approximation, which simplifies calculations for complex systems.\cite{rehrParameterfreeCalculationsXray2010b} FDMNES also supports the full-potential finite difference method, which avoids potential limitations from muffin-tin approximation by constructing a totally free potential shape. Such a non-muffin-tin effect is key to nanocluster simulations where the contribution from surface atoms is essential.\cite{kravtsovaAtomicElectronicStructure2014} OCEAN employs DFT calculations with the Bethe-Saltpeter equation approach, which includes excitonic effects and better reproduces the \ce{L_{3}}/\ce{L_{2}} ratios of light transition metals. \cite{pothsTheoreticalPerspectiveOperando2022} \textcolor{black}{We adopted local density approximation (LDA) exchange correlation functional for OCEAN computation and SCAN for VASP computations. }

\section{Results}
\subsection{Benchmarking of computed XANES}
\begin{figure}[H]
	\begin{center}
		\includegraphics[width=1\textwidth]{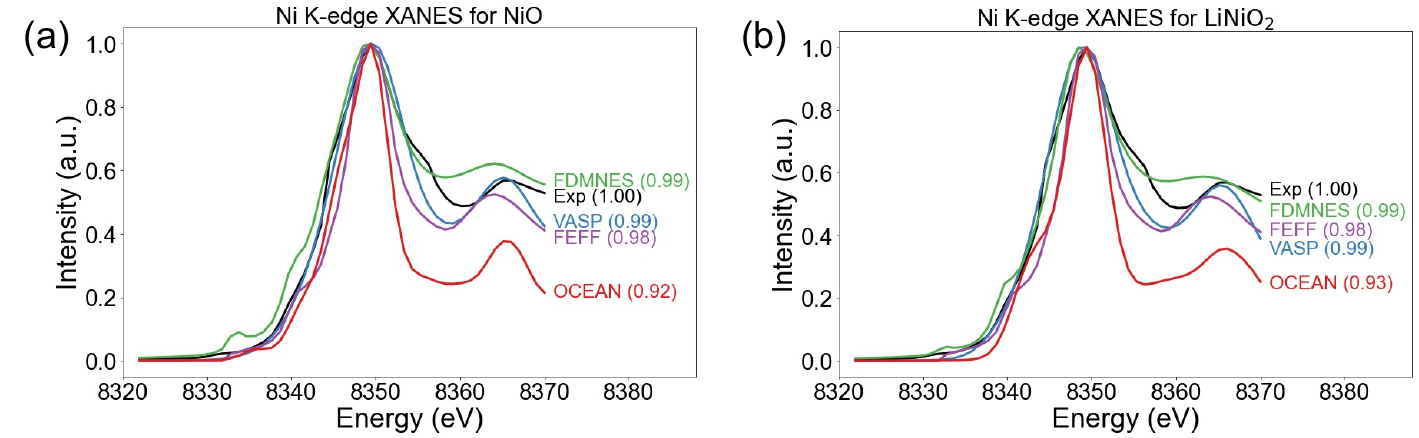}
				\caption{
		\label{fig:spec_benchmark}
				Ni K-edge XANES spectra for (a) NiO and (b) \ce{LiNiO2} computed using \textcolor{black}{VASP 6.1, FEFF 9.6, FDMNES, and OCEAN 2.5}, as well as measured experimentally. The values in brackets indicate the Pearson correlation between the computed and experimental spectra. A higher Pearson correlation indicates a higher similarity.}
		\end{center}
\end{figure}

Figure \ref{fig:spec_benchmark} compares the computed Ni K-edge XANES for NiO and \ce{LiNiO2} from different codes. The computed spectra were horizontally shifted to align the position of the maximum peak in the experimental spectrum. For both NiO and \ce{LiNiO2}, we found that all four codes - VASP, FEFF, FDMNES, and OCEAN - produced XANES that are in good agreement with experiments. Using the Pearson correlation coefficient as a metric, FEFF and VASP produced XANES that have the highest similarity with the experimental spectra.  Given that VASP was already used for structural relaxations, we adopted VASP for all subsequent XANES computations for ease of computational workflow. 

\subsection{Initial target analysis}

\begin{figure}
    \centering
	\begin{subfigure}[b]{0.47\textwidth}
	    \centering
	    \includegraphics[width=\textwidth]{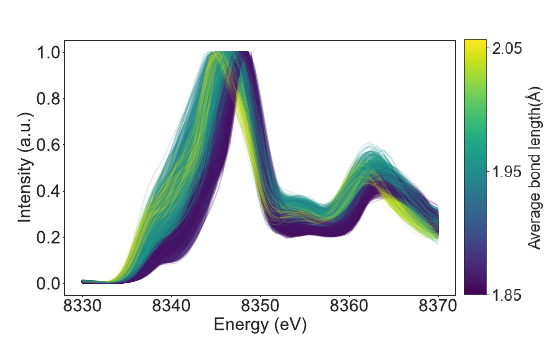}
	    \label{fig:spec_visalization_BL}
	    \subcaption{}
	\end{subfigure}
    \hfill
    \begin{subfigure}[b]{0.45\textwidth}
	    \centering
	    \includegraphics[width=\textwidth]{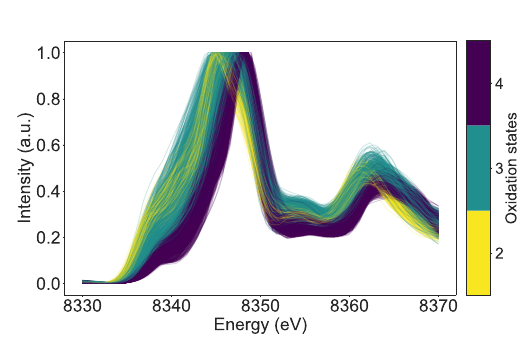}
	    \label{fig:spec_visalization_oxi}
	    \subcaption{}
	\end{subfigure}
    \hfill
	\begin{subfigure}[b]{0.4\textwidth}
	    \centering
	    \includegraphics[width=\textwidth]{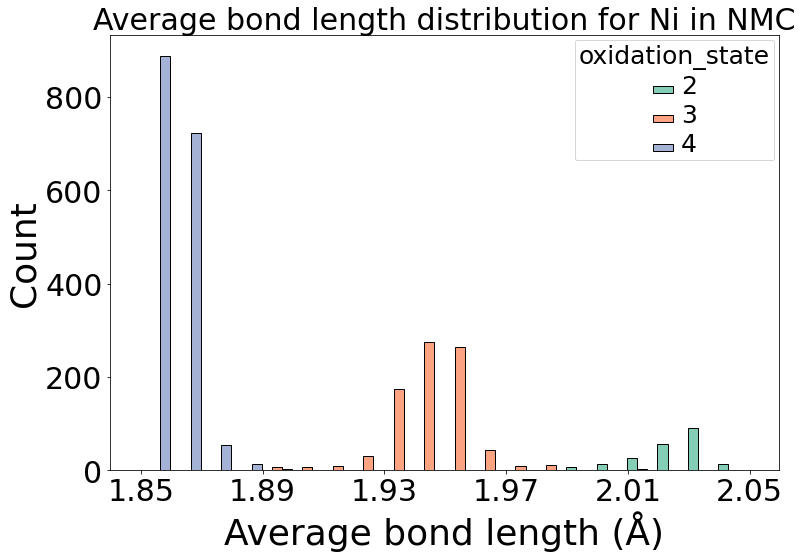}
	    \subcaption{}
	    \label{fig:BL_distribution}
	\end{subfigure}
	\caption{(a) Computed Ni K-edge XANES for NMC compounds, colored by average Ni-O bond length of absorbing atoms. (b) Computed Ni K-edge XANES for NMC compounds, colored by oxidation state of absorbing atoms. (c) Average bond length distribution for computed site-wise Ni K-edge XANES spectra.}
	\label{fig:spec_visualization}
\end{figure}

Figure \ref{fig:spec_visualization}(a) and (b) plot the computed site-wise XANES in the train dataset, colored by the average Ni-O bond lengths and formal Ni oxidation states, respectively. It may be observed that the spectra corresponding to shorter average bond lengths and higher Ni oxidation states tend to shift towards higher absorption energies. This is consistent with the fact that Ni atoms in higher oxidation states tend to have shorter bond lengths due to stronger electrostatic attraction, which in turn results in higher energy necessary to excite the outermost electrons in the absorbing atom. Figure \ref{fig:BL_distribution} plots the distribution of average bond lengths for different corresponding formal Ni oxidation states. It may be observed that the average Ni-O bond lengths for each formal Ni oxidation state are well separated.

\subsection{Oxidation states classification}

	\begin{figure}[H]
		\begin{center}
			\includegraphics[width=0.8\textwidth]{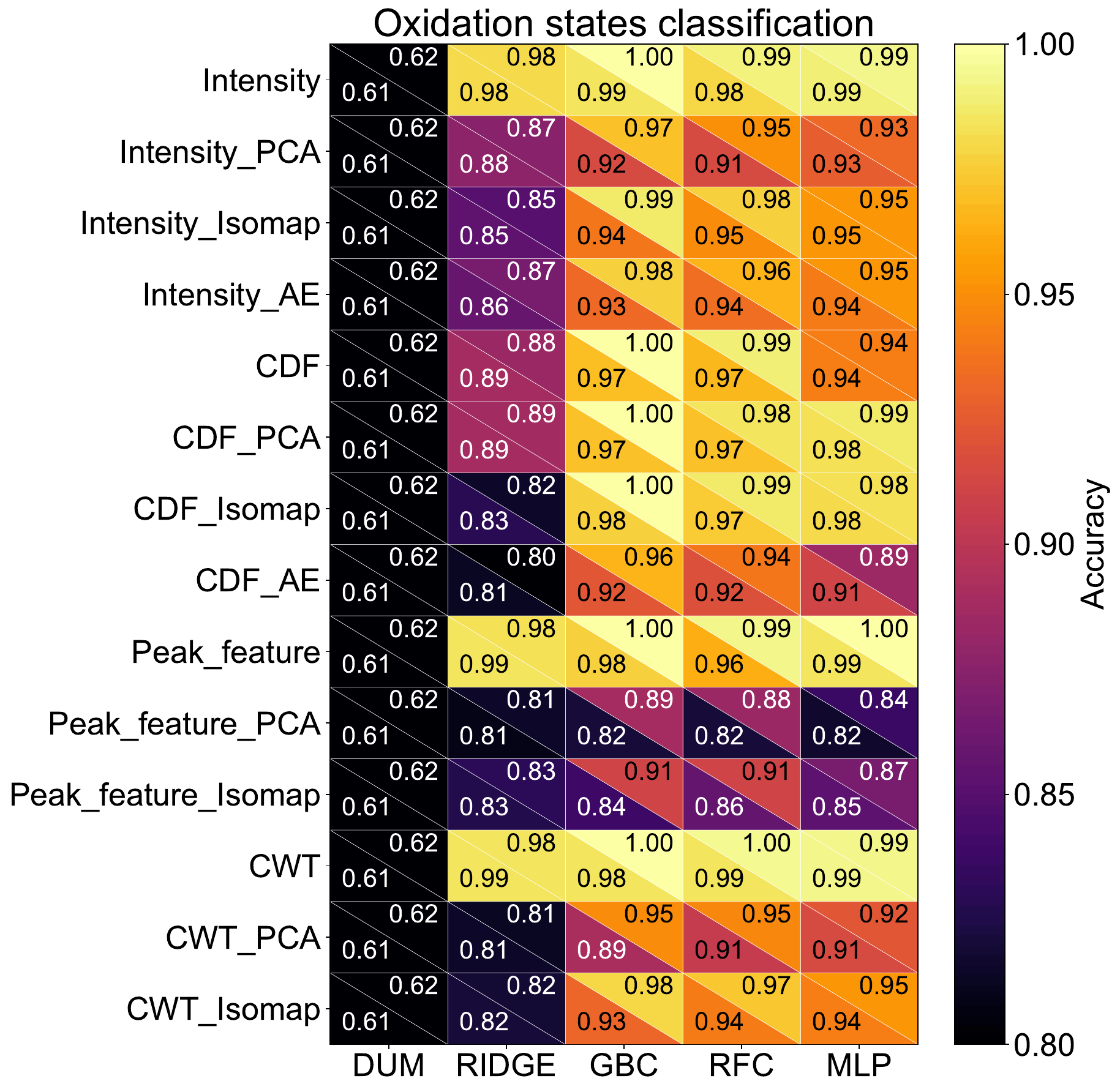}
			\caption{
					\label{fig:oxi_classification}
					Accuracy for oxidation states classification. The labels on the bottom represent various ML models for the ML tasks (DUM = dummy or baseline, RIDGE = ridge regression, GBC = gradient boosting classifier, RFC = random forest classifier, and MLP = multi-layer perceptron). The side labels are for different features (CDF = cululative distribution function, CWT = continuous wavelet transform, PCA = principle component analysis, AE = autoencoder). The upper right and lower left triangles within each cell represent train and test errors, respectively.}
		\end{center}
	\end{figure}
	
Figure \ref{fig:oxi_classification} compares the classification accuracy of different ML models and feature transformations in predicting the formal oxidation state from the computed site-wise XANES. All raw features (i.e., CWT, CDF, and peak feature) exert a similar level of accuracy as compared to the baseline input, intensity. This indicates featurization of the original spectrum preserves vital information to establish the relation between spectral shape and oxidation states. Generally, the train and test errors are similar, suggesting that there is little to no overfitting. The introduction of dimensionality reduction inevitably decreases the prediction accuracy. 

However, the features reduced using Isomap experienced less decrease than PCA and AE. This suggests the possibility that the linear relationship between spectroscopy and properties of interest is not well-established. Such a non-linear relationship is also implied by the inferior performance when using linear machine learning models such as ridge regression. CDF experiences a marginal decrease after dimensionality reduction, suggesting that CDF plus dimensionality reduction techniques can be an optimal combination for large-scale applications. 

\subsection{Average bond length regression}
The traditional way to obtain the average bond length around an absorbing atom is to perform fitting based on the EXAFS. However, in this study, we showcased that such information can be directly obtained from the XANES region of the spectra using ML without the EXAFS data. Although a similar schema was applied to this average bond length problem as to the oxidation states task, a regression problem that uses root mean squared error (RMSE) as the error metric can more appropriately describe the problem. 
	\begin{figure}[H]
		\begin{center}
			\includegraphics[width=0.8\textwidth]{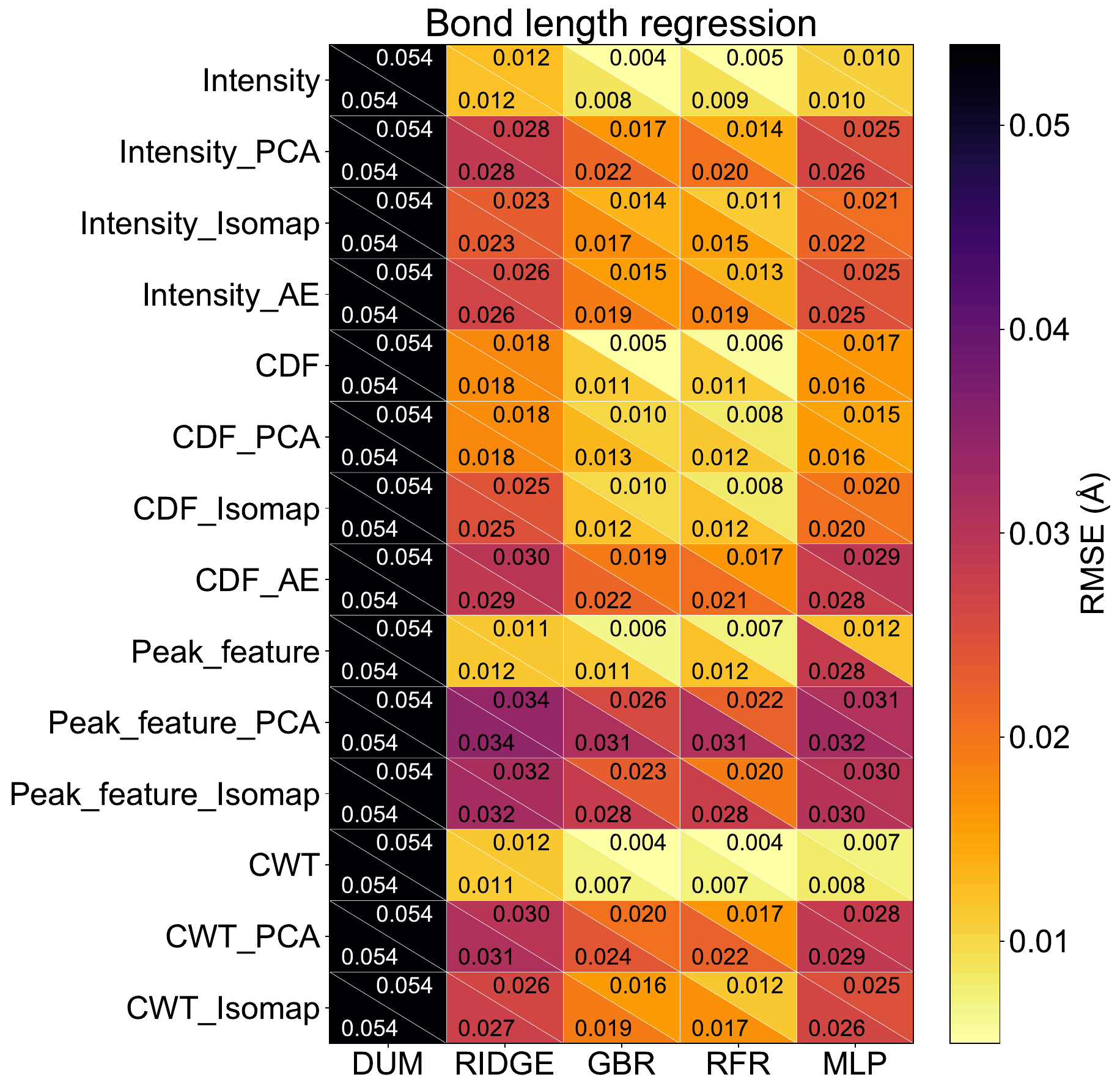}
			\caption{
					\label{fig:BL_regression}
					Root-mean-square errors for average bond length regression. The x label represents various ML models for the ML tasks and y labels are for different features, similar to Figure \ref{fig:oxi_classification}. The upper right and lower left triangles within each cell represent train and test errors, respectively.}
		\end{center}
	\end{figure}

The RMSEs for the train and test dataset are shown in Figure  \ref{fig:BL_regression}. Similar to the layout for the oxidation states classification, a lighter color in the heatmap represents a lower RMSE, hence a more accurate prediction. The major conclusions found in oxidation states classification align with those found in bond length regression, partially because the oxidation states and average bond length are highly correlated (as demonstrated in Figure \ref{fig:BL_distribution}). 
Because ensemble models (i.e., RFR and RFC) achieved the highest prediction accuracy for both regression and classification tasks, we decided to adopt random forest models for later analysis. 

\subsection {Inference on Experimental Data}
While the above results show that different featurization approaches (intensity, CDF, CWT) have similar performance when it comes to bond length and oxidation state inference on computed site-wise spectra, it is important to determine if these results hold when pre-trained ML models are applied towards data not used in training, especially experimentally-obtained spectra. In this section, we consider the performance of the featurization approaches on three experimental datasets. Different from the previous section where we trained models using only computational site-wise spectra, we used computational compound datasets for training purposes in this section.

Generally, all experimental spectra are site-averaged over all atoms of the same element. Therefore, compound or site-averaged spectra are needed for a direct comparison to experiment ones. The detailed procedures to collect experimental spectra and to construct compound datasets have been described in the Methods section. 

\begin{figure}[H]
		\begin{center}
			\includegraphics[width=1\textwidth]{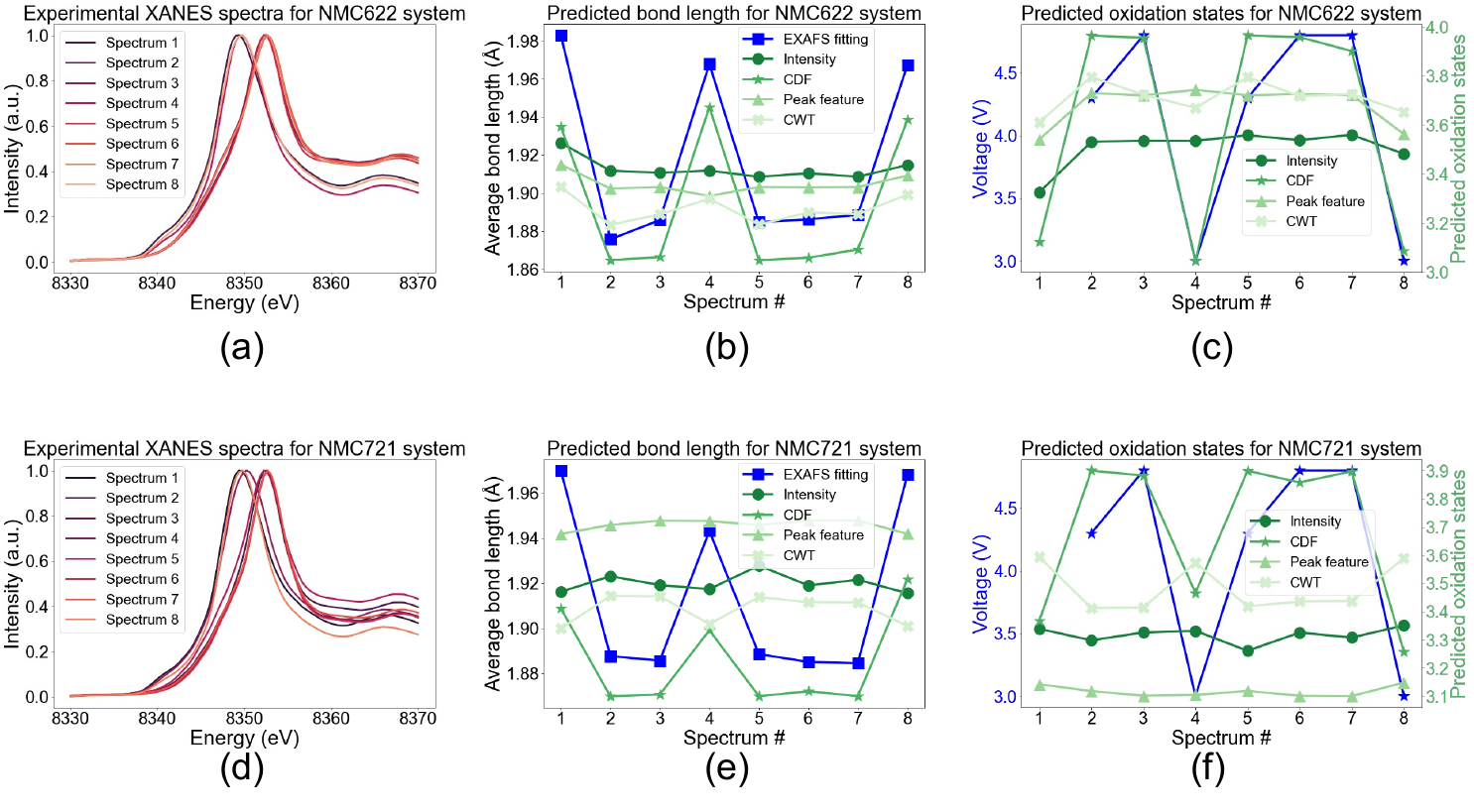}
			\caption{
					\label{fig:exp_validation}
					Performance of the trained random forest models on experimental NMC datasets. (a)-(c) show  NMC622 results and (d)-(f) show NMC721 results. (a, d)Ex-situ measurements of Ni K-edge XANES. (b, e) Average bond length from EXAFS analysis and predicted average bond length from trained models. (c, f) Voltage during battery cycling where ex-situ samples were taken for XAS measurements and predicted oxidation states. 
     The voltage for spectrum 1 was left blank because it corresponded to pristine sample. 
					}
		\end{center}
	\end{figure}

As suggested in Figure \ref{fig:exp_validation}(a), the experimental spectra for NMC622 samples form two distinct groups. Spectra 1, 4, and 8 form one group with lower peak energy, while the rest belong to the other group. Figure \ref{fig:exp_validation}(c) indicates the average Ni-O bond lengths from EXAFS fitting and predicted values using various features and the pre-trained RF models. The CDF feature achieves a quantitative agreement with our reference, the bond length from EXAFS fitting, with an average 2\% decrease compared to the values obtained from EXAFS fitting. This constant decrease may be attributed to the difference between DFT lattice parameters and experimental values, which is also around 2\%. A detailed lattice parameter comparison is available in Figure \ref{fig:lattice_comp}. Moreover, Figure \ref{fig:exp_validation}(c) displays both the corresponding voltage when measuring the XANES spectra and the inferred average oxidation states from compound spectra. While the ground truth average oxidation states are not known from these experimental data, we expect that the oxidation state will be higher for the higher voltage samples due to the removal of Li ions at high voltages. Moreover, as discussed before and shown in Figure 5c, shorter bond lengths should correspond to higher oxidation states. Although raw features including intensity, CDF, and CWT performed satisfactorily on the computed dataset (see Figure 6), only CDF gives the expected variations in the inferred oxidation state. The predicted oxidation states range from 3+ to 4+, which is also consistent with expectation. 

The results in Figure \ref{fig:exp_validation}(d)-(f) for NMC721 lead to similar observations as NMC622 systems, supporting the feasibility of using CDF to infer average bond length and oxidation states in unseen experimental data. Additionally, Figure \ref{fig:tsne} shows the t-SNE distributions for compound NMC622 spectra. While t-SNE distributions for intensity, peak feature, and CWT form a continuous band in 2D projections, the distribution for CDF is more separated, implying a higher possibility of distinguishing different oxidation states. 

\begin{figure}
    \begin{center}
    \includegraphics[width=0.95\textwidth]{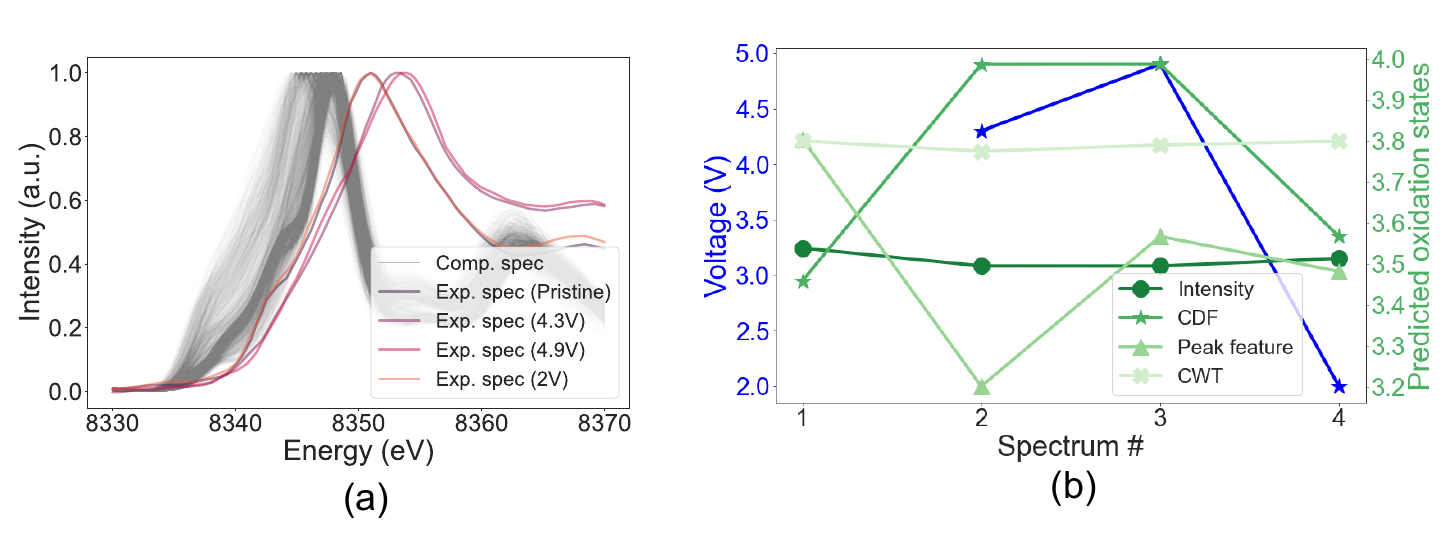}
        \caption{
        \label{fig:exp_validation_lirich}
            Performance of the trained random forest models on experimental Li-rich 0.5\ce{Li_{2}MnO_{3}}$\cdot$0.5\ce{LiMn_{0.5}Ni_{0.35}Co_{0.15}O_{3}} dataset\cite{atesHighRateLirich2015}. (a) Visualization of experimental spectra and computed spectra. (b) Relation between corresponding voltage and predicted oxidation states. The voltage for spectrum 1 was left blank because it corresponded to pristine sample. 
    }
    \end{center}
\end{figure}

An independent dataset of Ni K-edge XANES spectra for a Li-rich system was collected from Reference\cite{atesHighRateLirich2015} and replotted along with the computed site-wise spectra in Figure \ref{fig:exp_validation_lirich}(a). This chemical system, which consisted of 0.5\ce{Li_{2}MnO_{3}}$\cdot$0.5\ce{LiMn_{0.5}Ni_{0.35}Co_{0.15}O_{3}}, was similar to previous NMC systems but has a different transition metal ratio, as well as contains additional \ce{Li_{2}MnO_{3}}. Since the chemical system was different and there was no reference spectrum available in the computational dataset, no additional horizontal shift was applied, resulting in a several-eV shift between the computed and experimental spectra. However, CDF was still able to generate oxidation states that were expected for the measured voltage (Figure \ref{fig:exp_validation_lirich}(b)), while raw intensity, CWT, and peak feature produced unusable results. \textcolor{black}{Similar observations were also found from another independent dataset from Li et al. who investigated \ce{Li_{1.2}Mn_{0.6}Ni_{0.2}O_{2}} (LMNO) system\cite{li_improving_2022}. The corresponding results were shown in Figure \ref{fig:exp_validation_LMNO}. } Despite the challenges posed by a different chemical system and lack of precise energy alignment, CDF demonstrated its robustness as a featurization approach for applying models trained from computed spectra to real-world data. \textcolor{black}{Applying the model on chemistries outside of the training data gives qualitatively reasonable results, but for more quantitative inference, transfer learning with additional training data should be considered.}

\subsection{Model interpretability}
\begin{figure}[H]
		\begin{center}
			\includegraphics[width=1\textwidth]{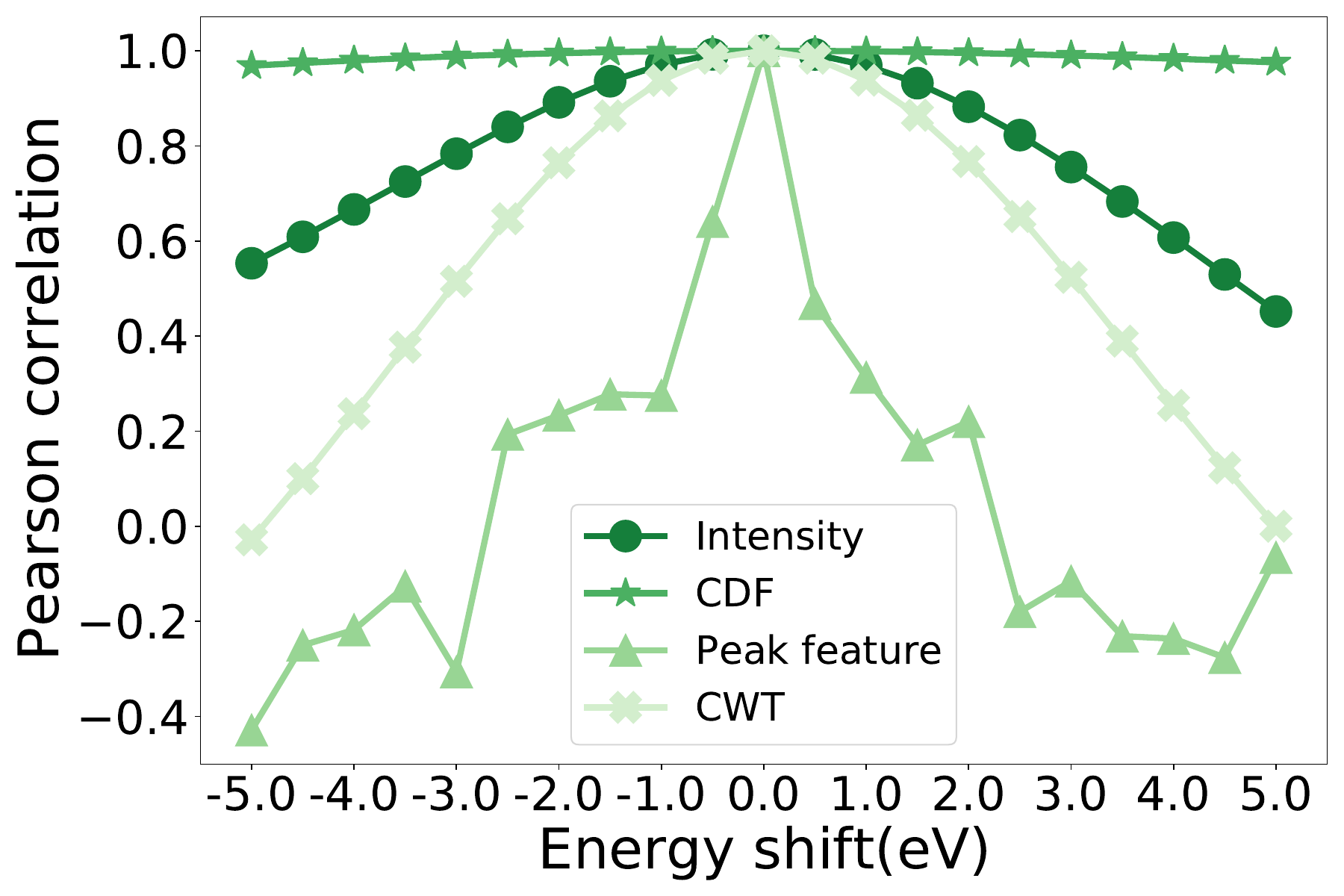}
			\caption{
					\label{fig:pearson}
					Pearson correlation between spectra with and without horizontal shift. 0 eV energy shift (no energy shift) was set as the reference to compute Pearson correlation for each feature.    
					}
		\end{center}
	\end{figure}

Energy alignment is a challenge for XAS researchers, as its causes can be complex and difficult to pinpoint. Factors such as sample preparation, XAS measurements, and post-processing may all contribute to the horizontal shift if the data are not well-processed/treated. 
Consequently, an energy adjustment is typically necessary when comparing two datasets. In this section, we investigated the tolerance to energy shift for the intensity, CDF, peak feature, and CWT. For each featurization approach, we calculated the Pearson correlation between the shifted spectra, with the Ni K-edge XANES spectrum for pristine NMC622 serving as the reference. As shown in Figure \ref{fig:pearson}, CDF demonstrated the least decrease in Pearson correlation when comparing shifted and non-shifted spectra, indicating its superior tolerance to energy shifts. Even in extreme cases such as 5 eV and -5 eV shift, the Pearson correlations for CDF were extremely high, at 0.97 and 0.98, respectively. In contrast, the baseline feature, intensity, experienced a much more drastic decrease (e.g., from 1.0 to 0.5 with a 5 eV shift). This finding explains why CDF worked best for the Li-rich dataset, which had several eV difference between the computed and experimental datasets. \textcolor{black}{To the best of our knowledge, a 5 eV shift after manual alignment accounts for most deviations between experiments and computations. The success of CDF is a milestone in tackling specific alignment issues in ML spectral fingerprinting, but accurate computational prediction, as well as careful experimental calibration, of absolute spectral energy position will still be important for benchmarking and evaluation of future simulations and ML models.}

\section{Conclusion}

In summary, this work addresses a crucial gap in data-driven approaches for XANES analysis through an in-depth benchmark for spectra featurization. In contrast to most previous studies that focused on spectral intensity, we investigated different ways to featurize the spectra and discovered that the CDF feature achieves a delicate balance between high prediction accuracy and excellent transferability. This excellent robustness is ascribed to its tolerance to energy shifts in spectra, which is critical when validating models trained on unseen experimental spectra. \textcolor{black}{While CDF appears to alleviate the need for precise energy alignment, the success of model trained on simulated spectra towards inference from experimental measurements still depends heavily on the accuracy of the simulated spectral shape.} Although this study focused on a specific technique (XANES) and a specific edge for a family of battery materials, the use of CDF may be generally advantageous for spectroscopic studies, as shown in recent work on quantitative metrics for comparing molecular spectra.\cite{seifertComputationalOptimalTransport2021,seifertComputationalOptimalTransport2022,seifertComputationalOptimalTransport2023}

Despite the continuous progress in both computational theory and computing power, the gap between experimental and computational spectroscopy remains. Such a gap has been impeding the broader application of ML models trained on computed datasets on experimental data, and the availability of experimental data with known ground truth is severely limited. The current study suggests that the use of CDF may be a useful strategy in bridging this gap and enabling ML models to harness the wealth of computed data while also making robust and accurate inference when applied to experimental data.

\section{Acknowledgement}

\textcolor{black}{We thank Deyu Lu and Matthew Carbone for helpful discussion on spectral simulation and machine learning approaches.} This work is supported by the U.S. Department of Energy (DOE) Office of Science Scientific User Facilities project titled “Integrated Platform for Multimodal Data Capture, Exploration and Discovery Driven by AI Tools”. We also acknowledge the support provided the Data Infrastructure Building Blocks (DIBBS) Local Spectroscopy Data Infrastructure (LSDI) project funded by National Science Foundation (NSF), under Award Number 1640899. M.K.Y.C. and Y.C. acknowledge the support from the BES SUFD Early Career award. M.J.D. was supported by the U. S. Department of Energy, Office of Basic Energy Sciences, Division of Chemical Sciences, Geosciences, and Biosciences operating under Contract Number DE-AC02-06CH11357. Work performed at the Center for Nanoscale Materials and the Advanced Photon Source, both U.S. Department of Energy Office of Science User Facilities, was supported by the U.S. DOE, Office of Basic Energy Sciences, under Contract No. DE-AC02-06CH11357. This research used resources of the Advanced Light Source, which is a DOE Office of Science User Facility under contract No. DE-AC02-05CH11231. We gratefully acknowledge the computing resources provided on Bebop, a high-performance computing cluster operated by the Laboratory Computing Resource Center at Argonne National Laboratory. We also acknowledge the computational resources provided by Triton Shared Computing Cluster (TSCC) and Expanse at University of California, San Diego. This research used resources of the National Energy Research Scientific Computing Center, a DOE Office of Science User Facility supported by the Office of Science of the U.S. Department of Energy under Contract No. DE-AC02-05CH11231.

\section{Author contributions}

M.K.Y.C., S.P.O., and Y.C proposed the concept. Y.C carried out the calculations and following analysis with the help of C.C., and M.J.D. C.J., W.Y., I.H., and G.L prepared the sample and performed XANES measurements. \textcolor{black}{D.M., D.A., and J.M.A developed AIMMDB and enabled data dissemination. }Y.C prepared the initial draft of the manuscript. All authors contributed to the discussion and revisions of the manuscript.

\clearpage

\bibliography{refs}
\newpage
\section{Supporting information}
\textcolor{black}{Detailed NMC structure distribution;spectra distribution;DFT functional comparison on lattice parameter; inference on LMNO dataset; and T-SNE distribution}
\newpage
\begin{figure}[H]
	\begin{center}
		\includegraphics[width=0.85\textwidth]{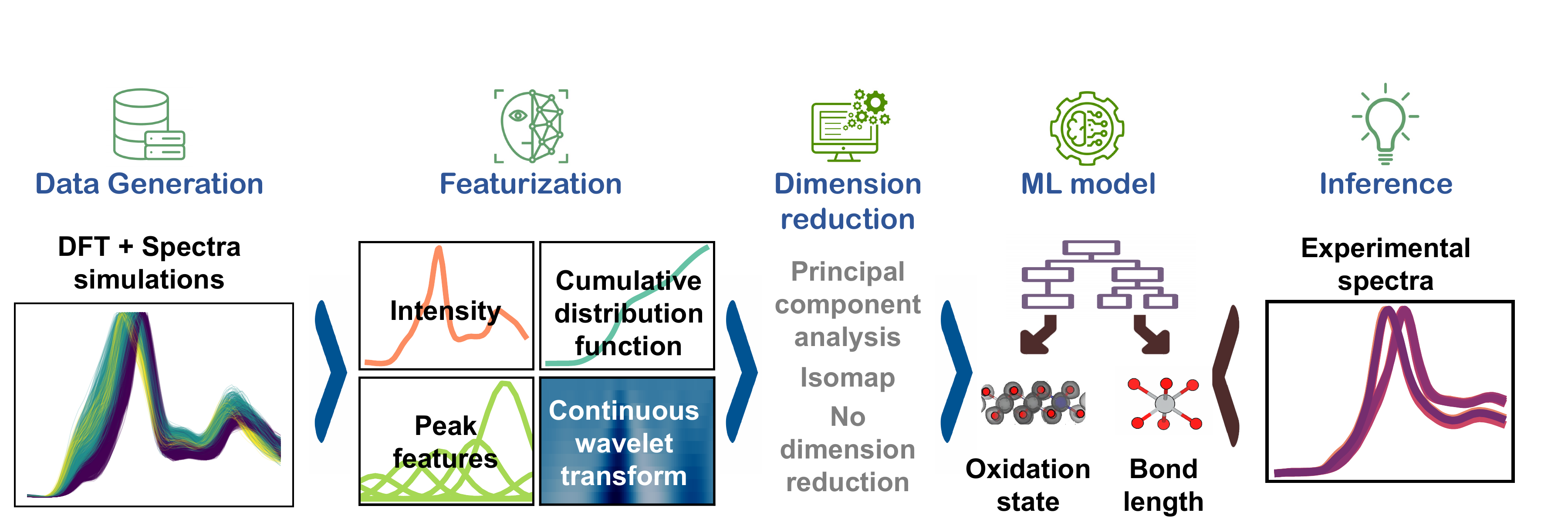}
	\end{center}
\end{figure}

\end{document}


\maketitle
\beginsupplement
\subsection{Structure distribution}
	\begin{figure}[H]
		\begin{center}
			\includegraphics[width=0.9\textwidth]{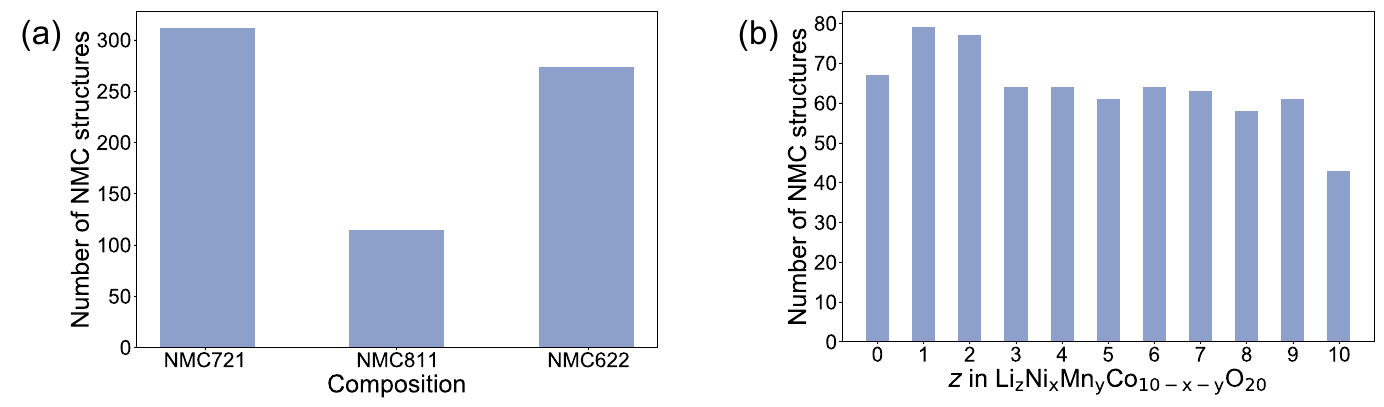}
			\caption{
				\label{fig:structure_distribution}
				NMC structure distribution for computed spectra with respect to (a)composition and (b) Li content. }
		\end{center}
	\end{figure}
	
\subsection{Spectra distribution}
	\begin{figure}[H]
		\begin{center}
			\includegraphics[width=0.9\textwidth]{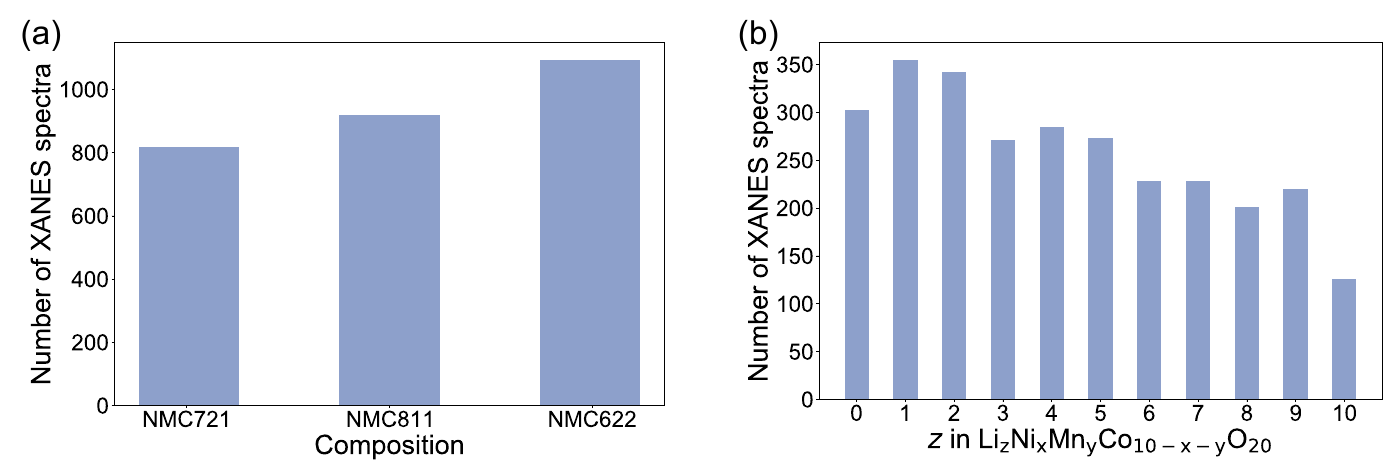}
			\caption{
				\label{fig:dataset_distribution}
				Spectra distribution for computed spectra with respect to (a)composition and (b) Li content. }
		\end{center}
	\end{figure}
\subsection{DFT functional comparison}
	\begin{figure}[H]
		\begin{center}
			\includegraphics[width=0.9\textwidth]{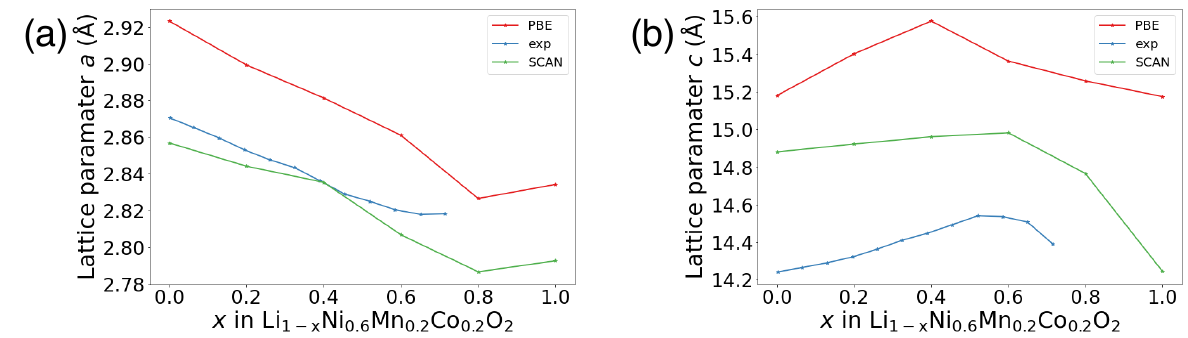}
			\caption{
				\label{fig:lattice_comp}
				Lattice parameter comparison between SCAN and PBE functional for DFT calculations. }
		\end{center}
	\end{figure}

\subsection{Inference on LMNO dataset}
 \begin{figure}[H]
		\begin{center}
			\includegraphics[width=0.9\textwidth]{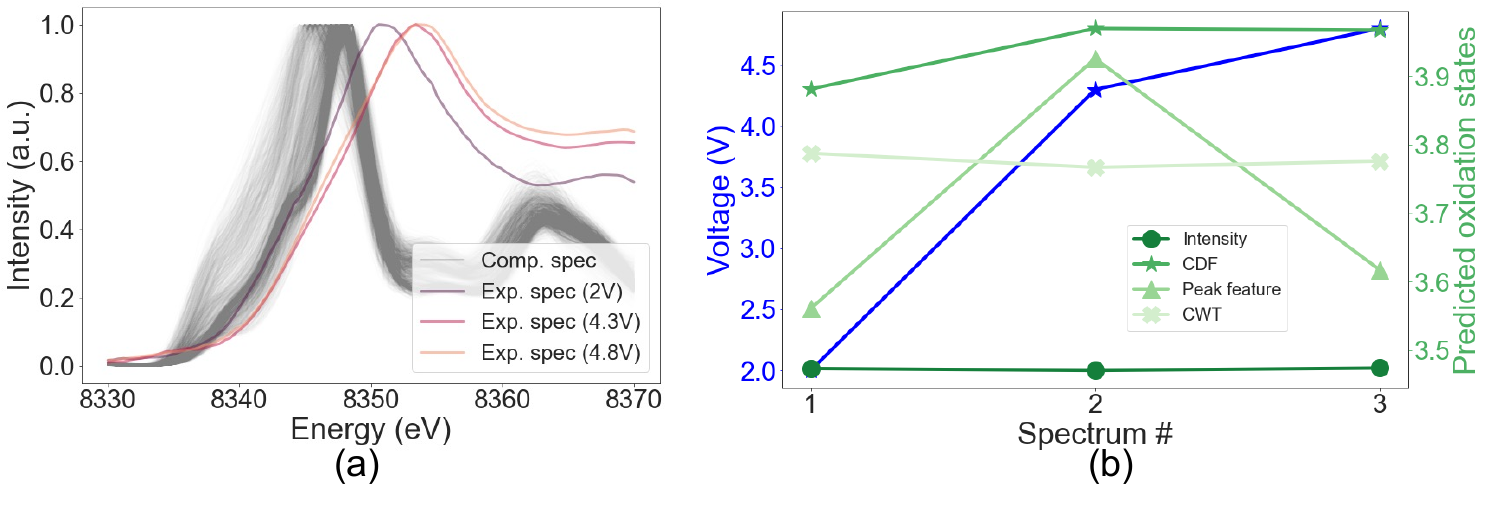}
			\caption{
				 Performance of the trained random forest models on experimental
     \ce{Li_{1.2}Mn_{0.6}Ni_{0.2}O_{2}} (LMNO) datasets.\cite{li_improving_2022} (a) Visualization of experimental spectra and computed spectra. (b) Relation between corresponding voltage and predicted oxidation states.  \label{fig:exp_validation_LMNO}}
				
		\end{center}
	\end{figure}

\subsection{T-SNE distribution}
    \begin{figure}[H]
		\begin{center}
			\includegraphics[width=0.9\textwidth]{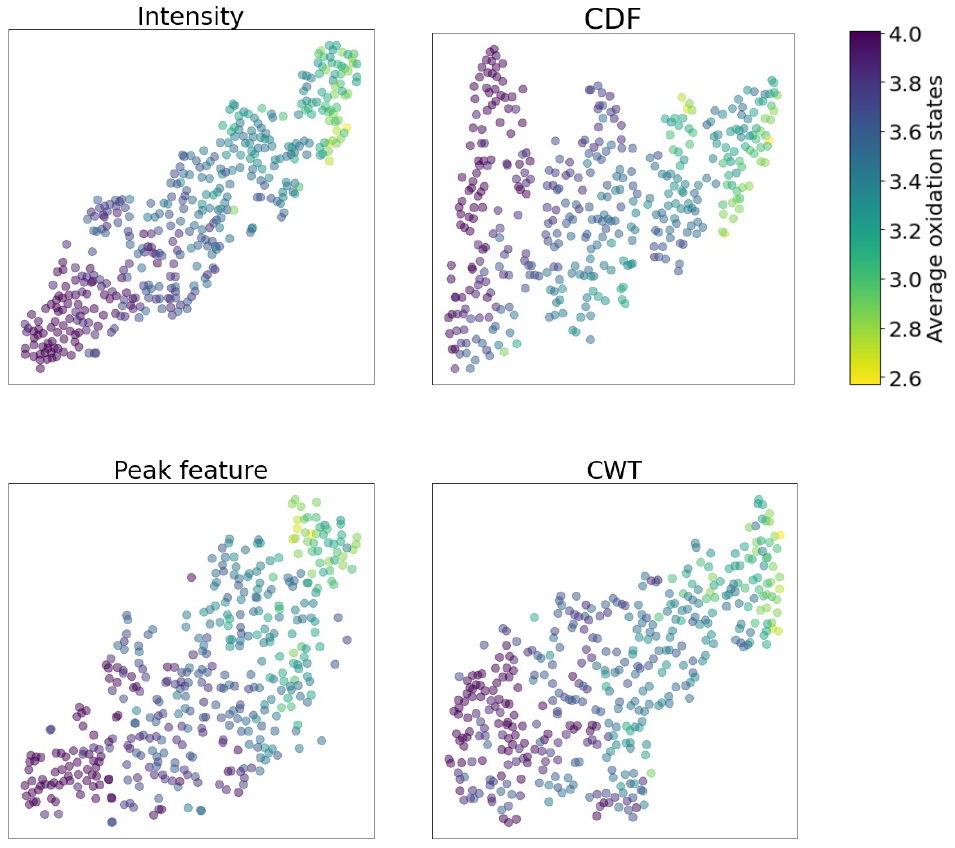}
			\caption{
				\label{fig:tsne}
				T-SNE distribution of  computed Ni K-edge XANES spectra. Each dot represents a site-averaged spectrum and is colored by its average oxidation states. 
    }
		\end{center}
	\end{figure}

\clearpage
\bibliography{refs}